\documentclass[showpacs,showkeys,preprint]{revtex4}
\usepackage{hyperref}
\usepackage{graphicx}
\usepackage{amssymb}
\usepackage{amsmath}

\begin{document}

\title{Using Muonic Hydrogen in Optical Spectroscopy Experiment to Detect Extra Dimensions}

\author{Feng Luo\footnote{
fluo@student.dlut.edu.cn} and Hongya Liu\footnote{
hyliu@dlut.edu.cn}}

\affiliation{Department of Physics, Dalian University of
Technology, Dalian, 116024, P. R. China}



\begin{abstract}
Considering that gravitational force might deviate from Newton's
inverse-square law (ISL) and become much stronger in small scale,
we propose a kind of optical spectroscopy experiment to detect
this possible deviation and take electronic, muonic and tauonic
hydrogen atoms as examples. This experiment might be used to
indirectly detect the deviation of ISL down to nanometer scale and
to explore the possibility of three extra dimensions in ADD's
model, while current direct gravity tests cannot break through
micron scale and go beyond two extra dimensions scenario.
\end{abstract}

\maketitle

\section{Introduction}

In the past few years, many physicists have dedicated themselves
to detecting possible deviation of Newton's inverse-square law
(ISL) at sub-millimeter scale. These activities are largely due to
a work done by N. Arkani-Hamed, S. Dimopoulos and G. Dvali (ADD)
in the late 1990s \cite{Arkani-Hamed}, \cite{Antoniadis},
\cite{Arkani}. In ADD's model, the long lasting hierarchy problem,
that is, the huge energy gap between Planck scale $M_{pl}\sim
10^{19}Gev$ and electroweak scale $m_{EW}\sim 10^{3}Gev$, was
tentatively solved by assuming that the fundamental energy scale
in nature is the $(4+n)$-dimensional Planck scale $M_{pl(4+n)}$,
where $n$ is the number of extra dimensions, while the
four-dimensional Planck scale $\sim 10^{19}Gev $ is just an
induced one. They further assumed that $M_{pl(4+n)}$ is around the
scale of $m_{EW}$, and the weakness of four dimensional
gravitational force is explained as that the originally strong
high dimensional one leaks into extra dimensions. Under the
assumption of $n$ equal-radii extra dimensions, they deduced that
the gravitational force would change from inverse-square law to
inverse-$(n+2)$ law if it is measured at a distance far smaller
than the radii of extra dimensions $R$, that is,
\begin{equation}
F\propto \frac{1}{r^{2}}\rightarrow \frac{1}{r^{n+2}},\text{for
}r\ll R. \label{eq1}
\end{equation}
(When $r\gg R$, the ordinary inverse-square law recovers.) In this
model, the radii of extra dimensions $R$ is determined by
$M_{pl(4+n)}$ as
\begin{equation}
R=\frac{1}{2\pi
}M_{pl(4+n)}^{-(1+\frac{2}{n})}G_{4}^{-\frac{1}{n}}, (c=1,\hbar
=1) \label{eq2}
\end{equation}
where $G_{4}$ is four-dimensional Newton's constant. For
$M_{pl(4+n)}=1Tev$,
\begin{equation}
R\approx \frac{1}{\pi }10^{-17+\frac{32}{n}}cm.  \label{eq3}
\end{equation}
If $n=1$, the radius of the extra dimension would be $R\sim
10^{12}m$. Clearly, this case is ruled out by planetary motion
observations. For $n=2$, $R\sim 10^{-1}mm$, but gravity
measurement had not been done at sub-millimeter scale by the time
of the presentation of ADD's model. After a few years' effort
(\cite{Long}, \cite{Price} and an extensive review in
\cite{Hoyle}), under the assumption of two equal-radii extra
dimensions, current experiment results require that $M_{pl(4+n)}$
should not be smaller than about $1Tev$, and the corresponding $R$
(notice Eq.(\ref{eq2})) should not be larger than about $100\mu
m$. (Of course, the probability of one large extra dimension with
several small extra dimensions can not be boldly excluded.) Then
what about three or more equal-radii extra dimensions? From
Eq.(\ref{eq3}), one can easily notice that $R$ decreases with the
increase of $n$. If we go one step further, that is, for three
equal-radii extra dimensions, we get $R\sim 10^{-7}cm$. However,
direct measurement of gravitational force (e.g. torsion pendulum
experiments) at nanometer scale is far beyond the ability of
current experiments since the background noises from such as
electrostatic, magnetic and Casimir forces would completely swamp
gravity effect.

Even if there is no motivation from ADD's proposal, the detection
of the possible deviation of ISL in small scale is still
meaningful, since there is no special reason to assume that ISL
should hold ranging from as large as solar system scale and as
small as nucleus scale. Moreover, we should point that except the
possible effect of extra dimensions, several other factors could
also cause the deviation of ISL in small scale \cite{Adelberger}.

At the same time, extensive researches aimed to explore extra
dimensions both in theories and experiments were done in the field
of accelerators \cite{Macesanu}. Constraints on the parameters of
extra dimensions, such as $M_{pl(4+n)}$, have also been given
\cite{Mele}, \cite{Luo}.

Considering current direct gravity measurement cannot be performed
down to smaller than tens of micron scale, we want to ask if we
could find other experimental methods to detect the deviation of
ISL at smaller than micron scale and even at nanometer scale?
Also, except current known experiments, such as the ones with
accelerators and torsion pendulums, if there could be other ways
in completely different fields to explore extra dimensions?

In this paper, we try to propose an experimental test in optical
spectroscopy to detect the possible deviation of ISL, and we use
electronic, muonic and tauonic hydrogen atoms as examples. This
proposal is based on a simple idea: if gravity deviates from ISL
and becomes much stronger in small scale, then in the filed of
optical spectroscopy, perhaps the originally safely neglected
gravity may need to be considered, since it might be large enough
to affect the optical spectrum of the atoms to the extent that we
could observe this effect in the lab. This kind of experiments
might help to set constraints of $M_{pl(4+n)}$ or the radii of
extra dimensions, and provide a way to further explore the
possibility of $n=3$ in ADD's model.

\section{A Possible Optical Spectroscopy Experiment to
Detect the Deviation of ISL}

First, take the ordinary hydrogen atom as an example.

Even if gravity would become stronger in small scale, it is still
very weak compared to electromagnetic force. So it is convenient
to treat the gravitational potential as a perturbation to
calculate the correction of atom's energy levels. Since $1Gev$
corresponds to $2.4\times 10^{23}Hz$ ($\Delta E=h\Delta \nu $),
and the frequencies of the spectrum correspond to the transitions
between different energy levels, then comparing the data measured
in experiment and the values calculated from Standard Model may
provide us some information about the deviation of ISL.

Notice that the correction for the ground state ($1s$) is much
larger than for the excited states (since Bohr's radius of ground
state is much smaller than the ones of excited states), then as to
the orders of magnitude, calculating the correction of the ground
state energy is enough if we just want to analyze the possible
changes of the spectrum generated from the transitions from some
excited states to the ground one.

For definition and simplicity, in the following calculation, we
employ ADD's model and assume there exists $n$ equal-radii
torus-shaped extra dimensions.

The leading correction comes from the Schrodinger term, that is,
for the ground state, we can use the simple perturbation theory of
quantum mechanics and obtain the first-order correction of energy
for hydrogen atom as
\begin{equation}
\Delta E=(\Psi _{100}, \hat{V} (r) \Psi_{100}), \label{eq4}
\end{equation}
where the wave function of ground state is $\Psi _{100}=\pi
^{-\frac{1}{2}}a^{-\frac{3}{2}}e^{-\frac{r}{a}}$, $a$ is Bohr's
radius of hydrogen atom $a=\frac{\hbar ^{2}}{me^{2}}$, $m$ is the
reduced mass $m=\frac{m_{l}m_{p}}{m_{l}+m_{p}}$, $m_{p}$ is the
mass of proton and $m_{l}$ is the mass of electron --- the
lightest lepton in nature. Here the gravitational potential is
expressed as \cite{Hoyle}, \cite{Kehagias}
\begin{equation}
\hat{V} (r)=\left\{
\begin{array}{cc}
-\frac{G_{(4+n)}m_{p}m_{l}}{r^{n+1}}, & r\ll R \\
-\frac{G_{4}m_{p}m_{l}}{r}(1+\alpha e^{-\frac{r}{\lambda }}), &
r\sim R
\\
-\frac{G_{4}m_{p}m_{l}}{r}, & r\gg R%
\end{array}%
\right. . \label{eq5}
\end{equation}
The $(4+n)$ Newton's constant is expressed as
\begin{equation}
G_{(4+n)}=\frac{2V_{n}G_{4}}{S_{n}},  \label{eq6}
\end{equation}
where $V_{n}=\left( 2\pi R\right) ^{n}$ is the volume of $n$ torus
and $S_{n}=\frac{2\pi ^{\frac{n+1}{2}}}{\Gamma (\frac{n+1}{2})}$
is its surface area ($\Gamma (n)=\int_{0}^{\infty
}e^{-x}x^{n-1}dx$). Notice that for $r\sim R$, we use the familiar
Yukawa potential. For the shape of extra dimensions we mentioned
above, we know that $\alpha =\frac{8n}{3}$ and $\lambda =R$. If
there is no correction of the gravitational potential from extra
dimensions, that is, for $n=0$, we get
\begin{equation}
\Delta E=-\frac{G_{4}m_{p}m_{l}}{a},  \label{eq7}
\end{equation}
which is just the correction from the consideration of the simple
semi-classical Bohr's hydrogen model.

For $n=2$, we get (notice Eq.(\ref{eq2}))
\begin{equation}
\Delta E\approx \frac{2m_{p}m_{l}M_{pl(4+n)}^{-4}}{\pi
a^{3}}\left[ \left( \ln (\frac{2r_{pro}}{a})+C\right) \right] ,
\label{eq8}
\end{equation}
where $C$ is the Euler-Mascheroni constant $C=0.577215\ldots $,
and $r_{pro}$ is of a value about proton's charge radius. For
$n=3$, we get
\begin{equation}
\Delta E\approx -\frac{4m_{p}m_{l}M_{pl(4+n)}^{-5}}{\pi
^{2}a^{3}r_{pro}}. \label{eq9}
\end{equation}
For $n>3$, the corresponding frequency corrections of spectrum are
so small that they are far beyond observation. Since with the
increase of $n$, $R$ becomes smaller, so the distance that the
deviated gravitational potential can act is shorter. Consequently,
the corrections decrease with the increase of $n$.

The parameter $M_{pl(4+n)}$ comes from the effect of extra
dimensions, while $r_{pro}$ is from the cutoff in the integration
of $\int_{r_{pro}}^{r\ll
R}\frac{e^{-\frac{2r}{a}}}{r^{n+1}}r^{2}dr$. Otherwise, for $n\ge
2$, this integration is divergent when $r\rightarrow 0$. Notice
that the proton is not a point like particle and the probability
that the electron appears in the interior of proton is extremely
small, so this cutoff is reasonable.

For hydrogen atom (e-p), if $M_{pl(4+n)}\sim 1Tev$, we get $\Delta
\nu \sim 10^{-8}Hz$ with $n=2$, and $\Delta \nu \sim 10^{-13}Hz$
with $n=3$. Although these corrections are much larger than the
one calculated from the exact ISL ($\Delta \nu \sim 10^{-24}Hz$),
it is still too small to be observed in current experiments.

What's worse, the calculation of the spectrum from the Standard
Model is not accurate \textquotedblleft enough\textquotedblright.\
For example, the experimental value of hyperfine structure of
hydrogen atom ($21cm$ line) is $1420.4057517667(9)MHz$
\cite{Hellwig}, while the most complete theoretical result
(considering reduced mass, QED, and hadronic effects) is
$1420.4051(8)MHz$ \cite{Sapirstein}. So, even this small effect
lies within the accuracy of the experiment, there is every reason
that it would be swamped by the small \textquotedblleft
inconsistency\textquotedblright\ of the data given by the Standard
Model and the observation. In other word, we must find some
\textquotedblleft large\textquotedblright\ correction that cannot
be simply explained out by the inconsistency of the accuracy of
the experimental data and Standard Model theoretical values.

Fortunately, nature gives us three generations of leptons. We can
treat muon and tau as \textquotedblleft heavy\textquotedblright\
electrons and do the similar calculation using muonic hydrogen
($\mu $-p) and tauonic hydrogen ($\tau $-p). However, we should
point out that for $\mu $-p and $\tau $-p, since the corresponding
Bohr's radii are far smaller than e-p, the influence from the
proton is much more important. (In fact, $\mu $-p has been used to
research nucleus for several decades \cite{Wu}, \cite{In}.) It is
even likely that the effects of gravity would be submerged by our
incomplete knowledge of proton's structure (about its charge and
gluon distribution etc.) and other uncertainties. Moreover, as to
the spectroscopy experiment itself, the measurement of the
transition energies we mentioned above may also not accurate
enough. Consequently, perhaps it is better to study the well
researched spectrum, such as the hyperfine structure, or to use
other atoms' spectrum to extract the possible information about
extra dimensions. In this sense, the calculation below may merely
a rough description about the orders of magnitude of the
corrections.

Anyway, if we use the simple perturbation theory, the main
correction is still from the Schrodinger term. From Eq.(\ref{eq8})
and Eq.(\ref{eq9}), we can easily find that $\Delta E$, or say,
$\Delta \nu  $ is relevant to the mass of lepton as $\Delta \nu
\propto m_{l}^{4}$, since $a=\frac{\hbar ^{2}}{me^{2}}$ and
$m\approx m_{l}$. Because the masses of muon and tau are
representatively about $200$ and $3000$ times of the mass of
electron, the corrections of the frequencies for $\mu $-p and
$\tau $-p are much larger.

For instance, if we choose $r_{pro}=10^{-13}cm$ and
$M_{pl(4+n)}=1Tev$, for $\mu $-p, we have $\Delta \nu \sim
10^{1}Hz$ with $n=2$ and $\Delta \nu \sim 10^{-4}Hz$ with $n=3$;
for $\tau $-p, $\Delta \nu \sim 10^{4}Hz$ with $n=2$ and $\Delta
\nu \sim 10^{0}Hz$ with $n=3$.

The adjustable parameter in the formula $\Delta E$ (also $\Delta
\nu$) is $M_{pl(4+n)}$. Consequently, in the field of optical
spectroscopy, comparing the theoretical calculations and
experimental data might help to set constraint of $M_{pl(4+n)}$.

\begin{figure}[!t]
\centering
\includegraphics[width=3.0in]{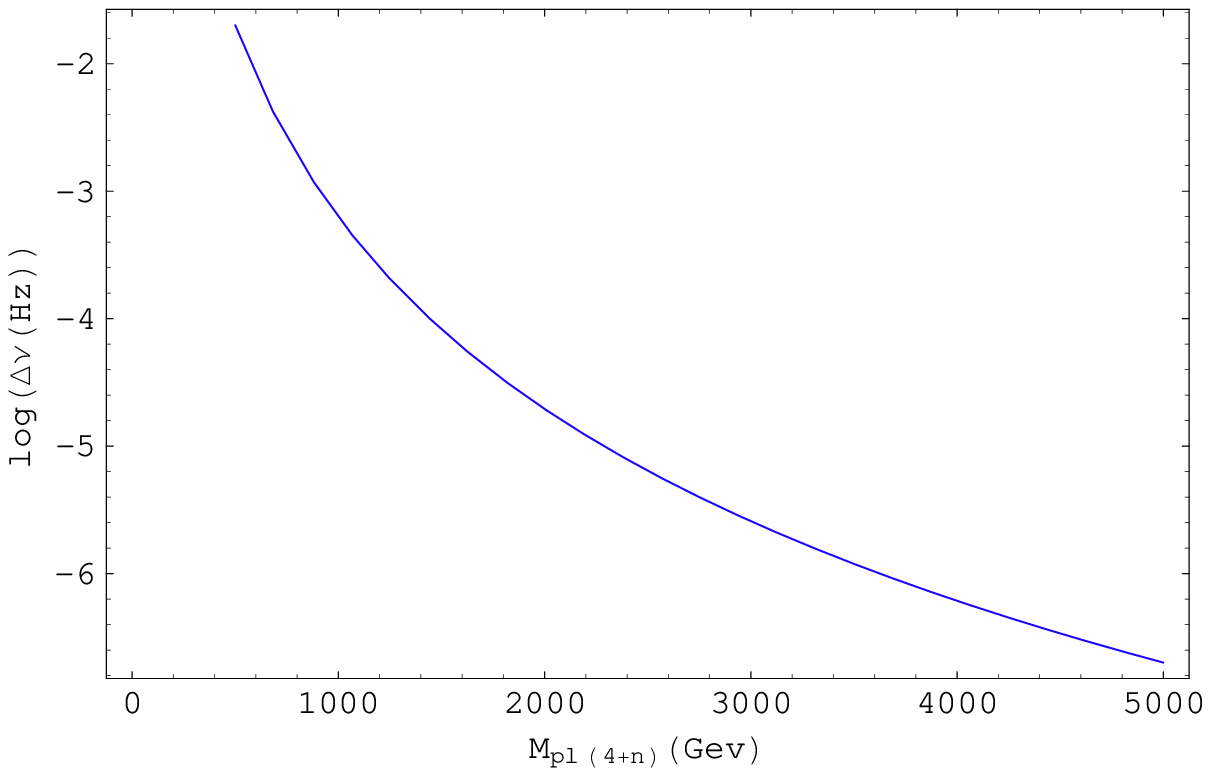}\\
\caption{$\mu $-p $(n=3,r_{pro}=10^{-13}cm)$. From Eq.(\ref{eq2}),
when $M_{pl(4+n)}$ ranges from $500Gev$ to $5000Gev$, $R$ ranges
from $5.3 nm$ to $0.1 nm$. (Note: the logarithm is $10$-based)}
\label{F2}
\end{figure}

It is easy to see the sensitivity of frequency dependent on
$M_{pl(4+n)}$ and the corresponding $R$ (notice Eq.(\ref{eq2})) in
the following figures. (Considering $\mu $-p is better researched,
we only draw figures for $\mu $-p.) From Fig.~\ref{F2}, when
$M_{pl(4+n)}$ varies one order of magnitude, $\Delta \nu $ varies
five orders of magnitude.

\begin{figure}[!t]
\centering
\includegraphics[width=3.0in]{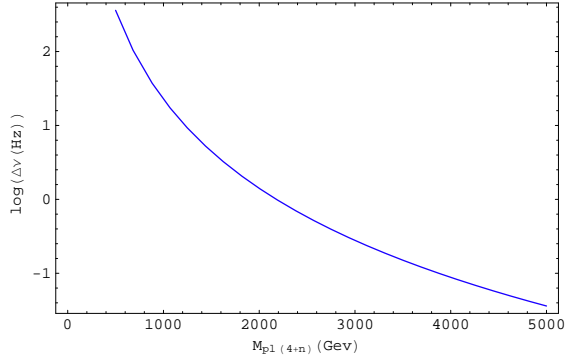}
\caption{$\mu $-p $(n=2,r_{pro}=10^{-13}cm )$. From
Eq.(\ref{eq2}), when $M_{pl(4+n)}$ ranges from $500Gev$ to
$5000Gev$, $R$ ranges from $1500 \mu m $ to $15 \mu m$. (Note: the
logarithm is $10$-based)} \label{F1}
\end{figure}

At last, we would like to point that just like $n=1$ case can be
excluded by the observations of celestial body motions, this
possibility may also be ruled out by optical spectroscopy
experiments. Since for e-p, $\Delta \nu \sim 10^{-1}Hz$; for $\mu
$-p, $\Delta \nu \sim 10^{5}Hz$; and for $\tau $-p, $\Delta \nu
\sim 10^{8}Hz$. These corrections may have already lain in the
accuracy of experiments but waiting for the calculation from
Standard Model.

\section{Conclusions and Discussions}

We propose an experimental test in optical spectroscopy to
indirectly detect the deviation of ISL. One of the most attractive
advantages is that this indirect experiment might be used to
detect the deviation of ISL down to nanometer scale, while current
direct gravitation experiments hardly break through micron scale.
Besides, within ADD's model and under the assumption of $n$
equal-radii torus-shaped extra dimensions, using simple
perturbation theory in quantum mechanics, we give the corrections
of transition frequencies for electronic, muonic and tauonic
hydrogen atoms due to the deviation of ISL. This experiment might
help to set constraints on $M_{pl(4+n)}$ and the corresponding
radii of extra dimensions $R$ since the corrections of frequencies
are sensitive to $M_{pl(4+n)}$. Moreover, this possible optical
spectroscopy experiment may be used to explore the possibility of
three equal radii extra dimensions ($n=3$), while current direct
gravity experiments go no further of $n=2$.

Finally, we have to say that although the accuracy needed for this
kind of experiment is a real challenge for experimentalists,
perhaps its feasibility is mainly limited by theoretical
calculation. Just like the story of the hyperfine structure of
hydrogen atom we mentioned above, the experimentalists may have to
wait for their fellow theorists providing enough accurate
theoretical values to compare with their experimental data.
Moreover, the atoms we suggest may be not proper to be used in
this kind of experiment. Since many problems, such as the lack of
knowledge of proton structure, would seriously impact the accuracy
of the calculation and swamp the small effect we suggest, using
other kinds of atoms and study different spectrums may be more
realizable. In this sense, our proposal just aims to inspire
people that perhaps there are methods beside the well-known ways
(e.g. accelerators and torsion pendulums) to explore the
mysterious extra dimensions.

\section*{Acknowledgments}

This work was supported by NSF (10273004) and NBRP (2003CB716300)
of P. R. China.

\end{document}